    \documentclass{emulateapj}
  \usepackage{amssymb}
  \usepackage{times}
  \usepackage{epsfig}
  \usepackage{graphics}
  \usepackage{natbib}
\def\simlt{\lower.5ex\hbox{$\; \buildrel < \over \sim \;$}}
\def\simgt{\lower.5ex\hbox{$\; \buildrel > \over \sim \;$}}

\def\sT{\sigma_{\rm T}}
\def\tauT{\tau_{\rm T}}

\def\be{\begin{equation}}
\def\ee{\end{equation}}
\def\beq{\begin{eqnarray}}
\def\eeq{\end{eqnarray}}

\def\g{\Gamma}
\def\hh{g}
\def\bb{b}
\def\Rph{R_\star}
\def\tph{t_\star}

\def\D{{\cal D}}
\def\thl{\tilde{\theta}}
\def\thc{\theta}
\def\mul{\tilde{\mu}}
\def\muc{\mu}
\def\nul{\tilde{\nu}}
\def\nuc{\nu}
\def\kl{\tilde{\kappa}_\nu}
\def\kc{\kappa_\nu}

\def\jl{\tilde{j}_\nu}

\def\Il{\tilde{I}_\nu}
\def\Ic{I_\nu}
\def\I{I}

  \def\Llab{L}
\def\Ilab{\tilde{I}}
\def\Flab{\tilde{F}}
\def\Omlab{\tilde{\Omega}}
\def\Om{\Omega}
\def\kkk{\kappa}
\def\Sl{\tilde{S}_\nu}
\def\Sc{S_\nu}
\def\S{S}

\def\tc{t}

\def\KK{{\cal K}}
\def\IN{{\cal I}}
\def\SN{{\cal S}}

\def\FNlab{\tilde{\cal F}}
\def\INlab{\tilde{\cal I}}
\def\SNlab{\tilde{\cal S}}

\def\QN{{\cal Q}}
\def\RN{{\cal R}}
\def\Q{Q}
\def\R{R}
\def\Qlab{\tilde{Q}}

  \def\jNlab{\tilde{\epsilon}}
  \def\jN{{\epsilon}}
\def\rin{r_{\rm in}}

  \def\fad{a}

\def\taur{\tau_{\rm ray}}

\def\tobs{t_{\rm obs}}

\def\dN{\dot{N}}

\def\x{x}

  \def\rls{r_\star}
  \def\thlls{\thl_\star}
  \def\mulls{\mul_\star}
  \def\mucls{\muc_\star}
  \def\thcls{\thc_\star}
  \def\phils{\phi_\star}
\def\Sec{Section}
\def\Secs{Sections}
\def\gsat{\g_{0}}

  \def\symb{,}

\newbox\grsign \setbox\grsign=\hbox{$>$} \newdimen\grdimen \grdimen=\ht\grsign
\newbox\simlessbox \newbox\simgreatbox \newbox\simpropbox
\setbox\simgreatbox=\hbox{\raise.5ex\hbox{$>$}\llap
     {\lower.5ex\hbox{$\sim$}}}\ht1=\grdimen\dp1=0pt
\setbox\simlessbox=\hbox{\raise.5ex\hbox{$<$}\llap
     {\lower.5ex\hbox{$\sim$}}}\ht2=\grdimen\dp2=0pt
\setbox\simpropbox=\hbox{\raise.5ex\hbox{$\propto$}\llap
     {\lower.5ex\hbox{$\sim$}}}\ht2=\grdimen\dp2=0pt
\def\simgt{\mathrel{\copy\simgreatbox}}
\def\simlt{\mathrel{\copy\simlessbox}}

\begin{document}

\title{Radiative transfer in ultra-relativistic outflows}

\author{Andrei M. Beloborodov\altaffilmark{1}}
\affil{Physics Department and Columbia Astrophysics Laboratory, 
Columbia University, 538  West 120th Street New York, NY 10027;
amb@phys.columbia.edu}
                                                                                
\altaffiltext{1}{Also at Astro-Space Center of Lebedev Physical
Institute, Profsojuznaja 84/32, Moscow 117810, Russia}

\begin{abstract}
Analytical and numerical solutions are obtained for
the equation of radiative transfer in ultra-relativistic opaque jets. 
The solution describes the initial trapping of radiation,
its adiabatic cooling, and the transition to transparency.
Two opposite regimes are examined: (1) Matter-dominated outflow. 
Surprisingly, radiation develops enormous anisotropy in the fluid frame 
before decoupling from the fluid. The radiation is strongly polarized.
(2) Radiation-dominated outflow. The transfer occurs as if radiation 
propagated in vacuum, preserving the angular distribution and the blackbody 
shape of the spectrum. The escaping radiation has a blackbody spectrum 
if (and only if) the outflow energy is dominated by radiation up to the 
photospheric radius.
\end{abstract}

\keywords{ radiative transfer --- relativistic processes --- scattering
--- gamma-ray burst: general }

%#########################################################################

\section{Introduction}

Powerful jets from compact objects can have significant optical depth
to scattering. The foremost example is gamma-ray bursts (GRBs).
They are emitted by hot ultra-relativistic outflows that remain opaque 
until they travel a large distance from the central engine.
Where the jet becomes transparent, the trapped radiation is released 
and contributes to the GRB. Its spectrum is expected to be nonthermal
because of dissipative processes in the subphotospheric region.

This ``photospheric emission'' is likely the main component of observed
GRBs. Recent work provides significant support for this picture. Three 
heating mechanisms have been proposed to shape the photospheric spectrum: 
(1) internal shocks,
(2) dissipation of magnetic energy and excited plasma waves
(Thompson 1994; Spruit, Daigne, \& Drenkhahn 2001; Ioka et al. 2007), and 
(3) collisional dissipation (Beloborodov 2010; hereafter B10). 
The latter mechanism is straightforward to model from first principles
and turns out to reproduce the canonical GRB spectrum with no fine-tuning 
of parameters (B10; Vurm, Beloborodov, \& Poutanen 2011). 

Modeling emission from opaque jets requires simulations of radiative 
transfer. Two methods have been developed for such simulations to date.
First, solving the kinetic equations for the electrons and photons that 
interact via Compton scattering (Pe'er \& Waxman 2005; Vurm et al. 2011).
Second, tracking a large number of photons that propagate and (randomly) 
scatter in the jet (Giannios 2006; B10). Pe'er (2008) used an analytic 
approach and Monte-Carlo simulations to study individual short pulses of 
photospheric emission.

None of these works attempted to use the standard transfer equation.
This approach is developed in the present paper. The extension of 
radiative transfer theory to relativistic outflows is straightforward 
(e.g. Castor 1972; Mihalas 1980). In \Sec~2, we write down the transfer 
equation that is well-behaved (and simplifies) in the ultra-relativistic 
regime. Then we solve this equation for isotropic (\Sec~3) and exact 
(\Sec~4) models of electron scattering. In parallel, we apply the 
independent Monte-Carlo technique and compare the results. 

In \Sec~5 we separately consider the case where radiation 
dominates the outflow energy up to the photosphere. 
This regime is of interest for GRBs with extremely low baryon loading,
as described by Paczy\'nski (1986) and Goodman (1986).
In contrast to their expectations, the transfer near the photosphere is 
not complicated and has a simple analytical solution.

%########################################################################

\section{Transfer equation in the ultra-relativistic regime}

\subsection{Formulation of the problem}

We are interested in outflows with Lorentz factors $\g\gg 1$ and velocities 
$\beta=v/c\rightarrow 1$. Below we consider radiative transfer in outflows 
that are steady and spherically symmetric. These assumptions are not 
restrictive in the ultra-relativistic regime, as discussed in \Sec~6, --- 
even strongly variable and beamed jets may be described by this model. 

The transfer equation is well-behaved in the limit $\beta\rightarrow 1$ 
when it is formulated for radiation intensity in the fluid frame 
(see Appendix~A). This frame is comoving with the outflow at any radius $r$. 
Hereafter intensity is denoted by $\Ic(r,\muc,\nuc)$ where $\nuc$ is the 
photon frequency, $\muc=\cos\thc$, and $\thc$ is the photon angle with 
respect to the radial direction. The quantities $\nuc$, $\muc$, and $\Ic$ 
are measured in the fluid frame. When $\beta\rightarrow 1$, the transfer 
equation~(\ref{eq:transfer_c1}) simplifies to
\beq
\nonumber
   \frac{\partial\Ic}{\partial \ln r}
  &=& -\left(1-\muc^2\right)\hh\,\frac{\partial\Ic}{\partial\muc}
      +(1-\muc\hh)\left(\frac{\partial\Ic}{\partial\ln\nuc}-3\Ic\right) \\
  &+& \tau_\nu\,\frac{(\Sc-\Ic)}{1+\muc},
\label{eq:transfer_c}
\eeq
where 
\be
\label{eq:h}
    \hh(r)\equiv 1-\frac{d\ln\g}{d\ln r}.
\ee
The quantity $\Sc$ appearing in equation~(\ref{eq:transfer_c}) is the source 
function in the fluid frame. It is determined by how radiation interacts 
with the outflow. For example, the simplest model of isotropic and coherent 
scattering gives $\Sc=(1/2)\int \Ic\,d\muc$ (e.g. Chandrasekhar 1960).

The quantity $\tau_\nu$ in equation~(\ref{eq:transfer_c}) 
  approximately represents the outflow optical depth (cf. Appendix~B).
It is defined as $\tau_\nu\equiv\kc r/\g$ where 
$\kc$ is the opacity in the fluid frame. In GRB jets, electron/positron 
scattering strongly dominates the opacity around the spectral peak of the 
burst. The scattering opacity is $\kc=\sigma n$ where $\sigma$ is the 
scattering cross section and $n$ is the proper $e^\pm$ density of the flow.
It may be expressed in terms of the rate of $e^\pm$ outflow through the 
sphere $4\pi r^2$ in the lab frame, $\dN_e=4\pi r^2\,n\,\g\,\beta c$. 
Then $\tau_\nu$ is given by
\be
\label{eq:tau}
  \tau_\nu=\frac{\sigma\,\dN_e}{4\pi r\,\g^2\,\beta c}.
\ee
When the outflow carries no positrons, $\dN_e$ remains constant with radius.
It also remains constant if positron creation is balanced by annihilation 
(this situation takes place in collisionally heated jets, see B10).

Besides the simpler form of the transfer equation, the ultra-relativistic
regime implies a principal change in the formulation of the transfer problem.
Radiation in the lab frame is strongly collimated and essentially all 
photons stream outward. Inward radial motion in the lab frame requires 
$\muc<-\beta$ in the fluid frame, which corresponds to a small solid angle 
$\Delta\Omega=2\pi(1-\beta)$.
At any given radius, only a small fraction ${\cal O}(\g^{-2})$ of all photons 
have $\mu<-\beta$.\footnote{
  This fraction equals $\Delta\Omega/4\pi\approx (2\g)^{-2}$ for a very 
  opaque flow, $\tau_\nu\rightarrow\infty$, where radiation isotropy is 
  maintained in the fluid frame. Deviations from isotropy that develop with 
  decreasing $\tau_\nu$ make this fraction even smaller, see \Sec~3.
}
As long as $\g\gg 1$, one can exclude the solid angle $\mu<-\beta$
from the transfer problem, i.e. neglect its contribution ${\cal O}(\g^{-2})$ 
when calculating the source function $\Sc$, and solve the problem in the 
domain $-\beta<\muc<1$. Then the outer boundary condition is not needed, 
as information cannot propagate from larger $r$ to small $r$. 

Thus, only an inner boundary condition should be specified for the 
ultra-relativistic transfer problem. If it is given at a sufficiently small 
radius $\rin$, the radiation may be assumed isotropic in the fluid frame
(as demonstrated by the solutions presented below,
radiation maintains isotropy where $\tau_\nu>100$).

Equation~(\ref{eq:transfer_c}) uses radius $r$ as an independent variable.
Alternatively, it can be written in terms of the comoving-observer time 
$t(r)$ which is related to $r$ by $d\tc=dr/\beta c\g$. Then the transfer 
problem takes the form of an initial-value problem. One can think of 
ultra-relativistic transfer as the {\it evolution} of intensity $\I(\muc)$, 
as seen by the comoving observer. This view is valid as long as radiation 
is not allowed to stream backward in time $t$ (i.e. backward in $r$).
The model is exact in the limit $\g\rightarrow\infty$ (then the entire 
region $-1<\mu<1$ is included in the transfer domain without violating 
causality). In practice, equation~(\ref{eq:transfer_c}) and the neglect 
of photons with $\mu<-\beta$ gives an excellent approximation when $\g>10$. 
The approximation is perfect for GRB jets.

\subsection{Transfer of energy}

The flow of radiation energy is described by the frequency-integrated
intensity,
\be
   \I(\muc,r)=\int_0^\infty \Ic\,d\nuc, \qquad
  \S(\muc,r)=\int_0^\infty \Sc\,d\nuc.
\ee
Integration of equation~(\ref{eq:transfer_c}) over $\nuc$ gives the equation 
for $I(\muc,r)$,
\be
\label{eq:trans}
   \frac{\partial\I}{\partial \ln r}
  =-\left(1-\muc^2\right)\,\hh\,\frac{\partial\I}{\partial\muc}
      - 4\,(1-\muc\hh)\,\I +\tau\,\frac{(\S-\I)}{1+\muc},
\ee
where $\tau(r)=\g^{-1}\kkk(r)r$ is defined using the effective opacity,
$\kkk=\I^{-1}\int \kc\,\Ic\,d\nuc$. 

Multiplying equation~(\ref{eq:trans}) by $1+\mu$ and integrating over $\mu$,
one gets
\be
\label{eq:en}
   \frac{d(\I_0+\I_1)}{d\ln r}
   =-4(\I_0+\I_1)+\hh\left(\I_0+2\I_1+\I_2\right)+\tau\,(\S_0-\I_0), 
\ee
where $\I_m$ ($m=0,1,2$) are the moments of intensity,
\be
\label{eq:mom}
   \I_m(r)=\frac{1}{2}\int_{-1}^1 \I(\muc,r)\,\muc^m\,d\muc,
\ee
and $\S_0(r)$ is the zero moment of the source function.
The quantities $4\pi\I_m$ are the components of the stress-energy 
tensor of radiation, and equation~(\ref{eq:en}) in essence expresses 
the first law of thermodynamics for radiation (e.g. Castor 1972).

In the case of coherent scattering (e.g. Thomson scattering by a plasma 
with a small Kompaneets' $y$-parameter) $\S_0=\I_0$. Then the last term in 
equation~(\ref{eq:en}) vanishes --- there is no heat exchange between the 
plasma and radiation, and the evolution of radiation with $r$ is adiabatic. 
Radiation does $PdV$ work and gradually gives away energy to the bulk 
kinetic energy of the outflow as it propagates to larger radii. This process 
of adiabatic cooling is easily described at large optical depths, $\tau\gg 1$, 
where radiation is nearly isotropic in the fluid frame, $\I_1=0$ and 
$\I_2=\I_0/3$. Then equation~(\ref{eq:en}) gives
\be
\label{eq:I0}
  \frac{d\ln\I_0}{d\ln r}=-4+\frac{4}{3}\,\hh.
\ee 
This equation reproduces the law of adiabatic cooling of radiation 
(adiabatic index $4/3$) in expanding volume. For example, if $\g(r)=const$ 
($\hh=1$, see eq.~\ref{eq:h}) volume expands as $r^{-2}$ and radiation energy 
density in the fluid frame $U=4\pi\I_0/c$ decreases as $r^{-8/3}$. 
The energy flux through the sphere $4\pi r^2$ measured in the lab frame 
decreases as $r^{-2/3}$. Note that equation~(\ref{eq:I0}) is valid only if 
radiation is isotropic in the fluid frame. Strong deviations from isotropy 
(which occur at optical depths $\tau\simlt 10$ as shown below) change the 
rate of adiabatic cooling. 

The fact that radiation does work on the outflow implies that the outflow 
accelerates and $\hh$ in equation~(\ref{eq:transfer_c}) is not an 
independent parameter of the transfer problem. It may be treated as a 
given fixed parameter only if the outflow inertia is large enough, so that 
it cannot be significantly accelerated by radiation. This condition reads 
$\rho c^2\gg U$ where $\rho$ is the rest-mass density of the outflow. 
This condition will be assumed in \Secs~3 and 4, where we solve the 
transfer equation with $\hh\approx1$ [i.e. $\g(r)\approx const$]. 
In \Sec~5, we consider the opposite regime and discuss the coupled 
dynamics of the outflow and radiation.

\subsection{Transfer of photon number}

The photon number intensity is described by the following quantity,
\be
   \IN(\muc,r)=\int_0^\infty \frac{\Ic}{h\nuc} \,d\nuc,
\ee
where $h$ is Planck constant. Integration of equation~(\ref{eq:transfer_c}) 
over $d\ln\nuc$ gives the equation for $\IN(\muc,r)$,
\be
\label{eq:trans_num}
   \frac{\partial\IN}{\partial \ln r}
  =-\left(1-\muc^2\right)\,\hh\,\frac{\partial\IN}{\partial\muc}
      - 3\,(1-\muc\hh)\,\I +\tau\,\frac{(\SN-\IN)}{1+\muc},
\ee
where $\SN=\int (\Sc/h\nuc)\,d\nuc$ and $\tau(r)=\g^{-1}\kkk(r)r$ is 
defined using $\kkk=\IN^{-1}\int \kc\,\IN\,d\nuc$.

Scattering conserves photon number, and the transfer equation is expected 
to give the corresponding conservation law. Multiplying 
equation~(\ref{eq:trans_num}) by $1+\mu$, integrating over $\muc$, 
and re-arranging terms, one gets
\be
\label{eq:num}
   \frac{d}{d\ln r}\,\ln\left[\g(\IN_1+\IN_0)\right]=-2,
\ee
where $\IN_m(r)$ are the moments of $\IN(\muc,r)$ and we used $\IN_0=\SN_0$ 
which is true for any scattering process. The quantity $4\pi\IN_1$ is the 
number flux of photons measured in the fluid frame, and $4\pi\IN_0/c$ is 
the number density of photons in the fluid frame. Lorentz transformation 
of the four-flux vector $4\pi(\IN_0,\IN_1,0,0)$ gives the radial photon 
flux measured in the lab frame, $\FNlab=4\pi\g(\IN_1+\beta\IN_0)$.
Equation~(\ref{eq:num}) in essence states $r^2\FNlab=const$
(with $\beta\rightarrow 1$) and expresses conservation of photon number.

%#########################################################################

\medskip

\section{Isotropic-scattering model}
\medskip

In this section, we solve the transfer problem assuming the simplest 
form of the interaction between radiation and the fluid: 
coherent isotropic scattering in the fluid frame. It gives a reasonable 
first approximation to Thomson scattering that is considered in \Sec~4. 
We consider here matter-dominated outflows --- the outflow is assumed to 
be massive enough, so that it can coast with $\g(r)\approx const$ (\Sec~2.2),
which corresponds to $\hh(r)\approx 1$ (eq.~\ref{eq:h}).

Then the energy transfer equation~(\ref{eq:trans}) reads, 
\be
  \label{eq:trans_iso}
  \frac{\partial\I}{\partial \ln r}
 =-\left(1-\muc^2\right)\,\frac{\partial\I}{\partial\muc}
      - 4(1-\muc)\,\I +\tau\,\frac{(\I_0-\I)}{1+\muc}.
\ee
Here we substituted the source function that describes isotropic scattering 
$S(\mu,r)=\I_0(r)$, where $\I_0$ is the zero-moment of intensity 
(eq.~\ref{eq:mom}). We will assume a constant cross section\footnote{
     This is a good approximation for the bulk of GRB photons. 
     The typical energy of observed photons is $\sim 1$~MeV. 
     They are emitted in the rest frame of the jet with energy 
     $\sim \g^{-1}$MeV, much smaller than $m_ec^2$. 
     Klein-Nishina corrections are small for such photons and the 
     scattering cross section is approximately independent of $\nu$.
}
$\sigma(\nu)=const$ and $\dN_e(r)=const$. Then equation~(\ref{eq:tau}) gives
\be
\label{eq:Rph}
 \tau(r)=\frac{\Rph}{r}, \qquad 
 \Rph=\frac{\sigma\,\dN_e}{4\pi\,\g^2\,\beta c}.
\ee

Transfer of photon number is described by equation similar to 
equation~(\ref{eq:trans_iso}) where $\I$ is replaced by $\IN$ and the 
numerical coefficient $-4$ in the second term on the right-hand side is 
replaced by $-3$ (cf.~eq.~\ref{eq:trans_num}).

\subsection{Integration of transfer equation}

Equation~(\ref{eq:trans_iso}) gives the expression for 
$\partial\I/\partial \ln r$ in terms of $\I$. Direct integration in $\ln r$ 
immediately yields the solution for $\I(\muc,r)$. Our numerical integration 
starts at $\rin=3\times 10^{-3}\Rph$ and takes the isotropic 
$\I(\muc,\rin)=const$ as the boundary condition. We use a uniform grid in 
$\thc$ and $\ln r$ of size $300\times 10^5$. With a simplest integrator 
--- Runge-Kutta scheme of fourth order --- the grid gives excellent 
accuracy of $\sim 0.1$\% (we have checked this by varying the grid).
Two more details of numerical integration are worth mentioning:

(1) At one boundary of the computational domain $\muc\rightarrow -1$
and the transfer equation gives $(\S-\I)/\I\rightarrow 0$.
This requires $\I=\S=\I_0$ at $\mu=-1$.
Note that the optical depth $\Delta\tau_{\rm ray}$ passed along the ray 
in one step $\Delta\ln r$ depends on $\muc$: 
$\Delta\tau_{\rm ray}(\muc)=\tau(r)\,\Delta\ln r/(1+\muc)$. 
Numerical integration is possible only if $\Delta\tau_{\rm ray}<1$,
which is violated close to the boundary $\mu=-1$.
However, in this region $\taur=\tau/(1+\muc)\gg 1$ simply enforces 
$\I\approx\S$. In the process of integration, we set $\I=\S$ wherever 
$\Delta\tau_{\rm ray}(\muc)>0.7$.

(2) The transfer equation contains the term 
$(1-\muc^2)\,\partial\I/\partial\muc=\sin\thc\,\partial\I/\partial\thc$.
We use a grid $\thc_i$ ($i=0,...,n$), where $\thc_0=0$
and $\thc_n=\pi$. The term $\sin\thc\,\partial\I/\partial\thc$ 
is not needed at $\theta_0$ and $\theta_n$ (it vanishes).
For all other $\theta_i$ we evaluate this term using 
$\partial\I/\partial\thc=(I_{i+1}-\I_{i-1})/(\thc_{i+1}-\thc_{i-1})$.

The transfer equation~(\ref{eq:trans_iso}) has no free parameters 
and the solution is unique. The result is shown in Figure~1. 
The striking feature is the strong beaming of the radiation field in 
the fluid frame, even at large optical depths $\tau\sim 10$. 
Beaming may be described by the ratio of intensities at $\mu=1$ and 
$\mu=-1$: $\bb(r)\equiv\I(1,r)/\I(-1,r)$. This quantity is shown in Figure~2.
It significantly deviates from unity starting at $\tau\sim 10$.
In the zone of $\tau\ll 1$, $\I(1,r)=const$ and $\I(-1,r)=\I_0\propto r^{-2}$.
Therefore $\bb(r)\propto r^2$ at $r\gg\Rph$.  

%%%%%%%%%%%%%%%%%%%%%%%%%%%%%%%%%%%%%%%%%%
\begin{figure}
% \label{fig:I0}
% \begin{center}
% \epsscale{1.2}
\epsscale{1.135}
\plotone{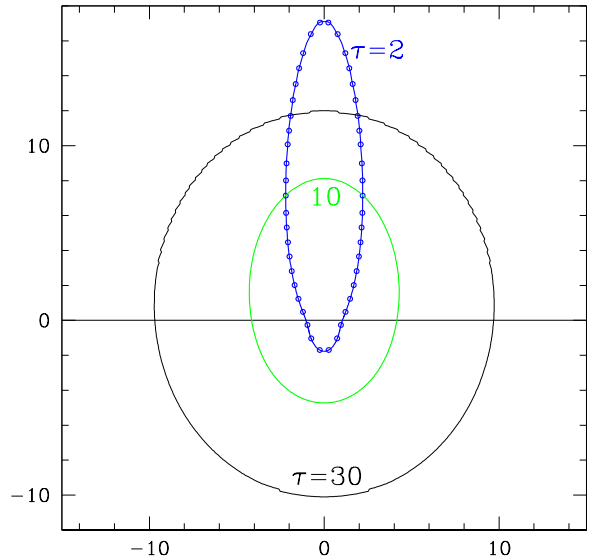}
\caption{
Angular distribution of radiation intensity in the fluid frame at three 
radii: $r/\Rph\approx 0.03$, 0.1 and 0.5, which correspond to optical depths 
$\tau\approx 30$, 10, and 2, respectively. For a better comparison, we plot 
$(r/\Rph)^2\I$, where the factor $(r/\Rph)^2$ compensates for the photon 
dilution due to expansion. The overall normalization of the transfer 
solution is chosen so that $\I(\muc,r)=(r/\Rph)^{-8/3}=\tau^{8/3}$ at 
radii $r<0.01\Rph$ where radiation is nearly isotropic ($\I$ does not 
depend on $\muc$) and follows the adiabatic cooling law $\I=\tau^{8/3}$.
Open circles show the intensity at $\tau=2$ that is obtained by the 
Monte-Carlo code (\Sec~3.3).
}
% \end{center}
\end{figure}
%%%%%%%%%%%%%%%%%%%%%%%%%%%%%%%%%%%%%%%%%%
%%%%%%%%%%%%%%%%%%%%%%%%%%%%%%%%%%%%%%%%%%
\begin{figure}
% \label{fig:I0}
% \begin{center}
% \epsscale{1.2}
\epsscale{1.1}
\plotone{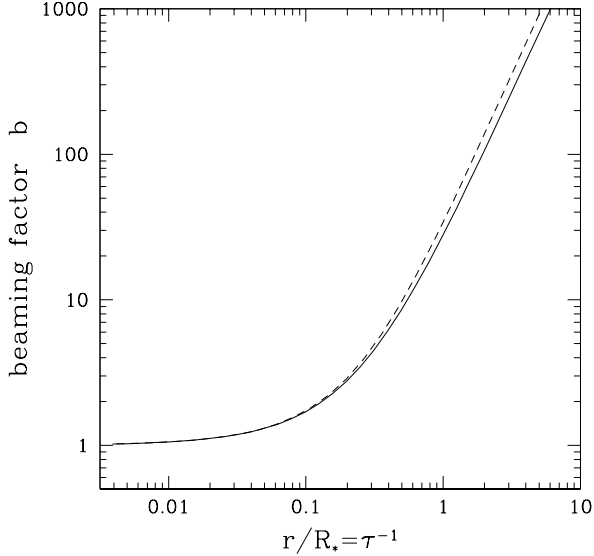}
\caption{
Beaming factor $\bb(r)=\I(1,r)/\I(-1,r)$ in the fluid frame.
Dashed curve shows the model with isotropic coherent scattering.
Solid curve shows the model with Thomson scattering in a cold outflow
(\Sec~4.1).  
}
% \end{center}
\end{figure}
%%%%%%%%%%%%%%%%%%%%%%%%%%%%%%%%%%%%%%%%%%
% \medskip

\subsection{Adiabatic cooling}

To examine adiabatic cooling, one should consider the energy
flux of radiation in the lab frame $\Flab=4\pi\Ilab_1$, 
where $\Ilab_1=\g^2[\beta(\I_0+\I_2)+(1+\beta^2)\I_1]$
is the first moment of intensity in the lab frame.\footnote{ 
    $4\pi\I_m$ are the components of the stress-energy tensor of 
    radiation: $T^{00}=4\pi\I_0$, $T^{01}=4\pi\I_1$, and $T^{11}=4\pi\I_2$, 
    where index 0 in $T^{\mu\nu}$ corresponds to the time coordinate and 
    index 1 corresponds to the spatial coordinate in the radial direction.
    Tensor transformation from the fluid frame to the lab frame reads
    $\tilde{T}^{\mu\nu}=\Lambda^{\mu}_{\sigma}\Lambda^\nu_\rho T^{\sigma\rho}$,
    where $\Lambda^0_0=\Lambda^1_1=\g$ and $\Lambda^0_1=\Lambda^1_0=\g\beta$.
    It gives $\tilde{T}^{01}=\g^2[\beta(T^{00}+T^{11})+(1+\beta^2)T^{10}]$.}
Here we cannot take the formal limit $\g\rightarrow\infty$, as the 
transformation between the lab frame and the fluid frame is not well defined 
in this limit. The total luminosity of radiation measured in the lab frame is
\be
   \Llab(r)=4\pi r^2\Flab
           =(4\pi)^2\,r^2\,\g^2\,\left[\beta\left(\I_0+\I_2\right)
                                      +\left(1+\beta^2\right)\I_1\right].
\ee
Freely streaming radiation would have $\Llab(r)=const$. Interaction with 
the outflow results in adiabatic cooling and $\Llab(r)$ decreases with $r$.

The adiabatic cooling factor is defined by 
$\fad(\rin\rightarrow r)=\Llab(r)/\Llab(\rin)$.
We chose $\rin\ll\Rph$ where radiation is nearly isotropic; then
$\I_1(\rin)=0$ and $\I_2(\rin)=\I_0(\rin)/3$. This gives,
\be
  \fad(\rin\rightarrow r)\equiv\frac{\Llab(r)}{\Llab(\rin)}
        =\frac{3}{4}\,\frac{r^2}{\rin^2}\,
         \frac{\beta\left(\I_0+\I_2\right)+\left(1+\beta^2\right)\I_1}
              {\I_0(\rin)}.
\ee
This equation is well-behaved in the limit $\beta\rightarrow 1$.
Adiabatic cooling is controlled by $\I_0(r)$, $\I_1(r)$, and $\I_2(r)$, 
which we know from the solution of the transfer equation. Figure~3 shows 
the resulting $\fad$. In the deep subphotospheric region (where radiation 
is approximately isotropic in the fluid frame) $\fad=(r/\rin)^{-2/3}$ as 
expected. A deviation from this law develops at $\tau\sim 10$ (see
also Monte-Carlo simulations in Pe'er 2008).
In the region $\tau\ll 1$ ($r\gg\Rph$) most of radiation streams
freely and experiences no adiabatic cooling.

The net effect of adiabatic cooling on the escaping radiation is 
described by
\be
   \fad(\rin\rightarrow\infty)=2\,\left(\frac{\rin}{\Rph}\right)^{2/3}
                              =2\,\tau_{\rm in}^{-2/3}.
\ee

%%%%%%%%%%%%%%%%%%%%%%%%%%%%%%%%%%%%%%%%%%
\begin{figure}
% \label{fig:I0}
% \begin{center}
% \epsscale{1.2}
\epsscale{1.1}
\plotone{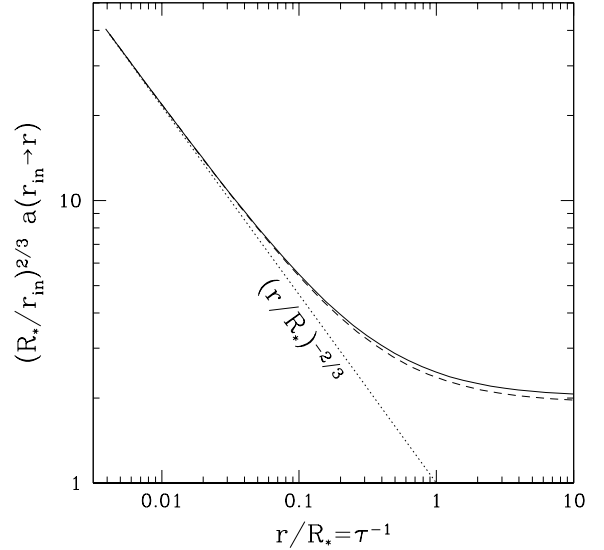}
\caption{
Adiabatic cooling factor $\fad(\rin\rightarrow r)$. 
We multiplied $\fad$ by $(\Rph/\rin)^{2/3}$ in this plot; 
this combination does not depend on $\rin$ as long as $\rin\ll\Rph$. 
Dotted line shows the thermodynamic result $\fad=(r/\Rph)^{-2/3}$, 
which is valid for approximately isotropic radiation. 
The actual cooling factor obtained from the transfer solution is shown 
by the solid and dashed curves.
Dashed curve: model with isotropic coherent scattering.
Solid curve: model with Thomson scattering in a cold outflow (\Sec~4.1).
}
% \end{center}
\end{figure}
%%%%%%%%%%%%%%%%%%%%%%%%%%%%%%%%%%%%%%%%%%

\subsection{Monte-Carlo simulation}

As an independent check of the results presented above, we solved the 
same transfer problem using the Monte-Carlo method. The numerical code is 
described in B10. In this section, we use its simplest version that assumes 
coherent isotropic scattering in the fluid frame.

The Monte-Carlo method is fundamentally different from solving the transfer 
equation. It operates with individual photons that are injected at  
$\rin\ll \Rph$. The simulation assumes a finite $\g$ and is performed in the 
lab frame. It tracks the propagation and random scattering of a large number 
of injected photons and accumulates their statistics at different radii. 
These statistics are used to reconstruct the angular distribution of 
radiation intensity in the fluid frame.
We ran the Monte-Carlo simulation for outflows with $\g=20$ and $\g=600$.
The results were identical, again confirming that the transfer does 
not depend on $\g$ as long as $\g\gg 1$ and $\g(r)=const$.

To reconstruct the radiation intensity at a given radius $r$ from the 
Monte-Carlo simulation, we calculate the angular distribution of luminosity
passing through the sphere of radius $r$ in the lab frame, $d\Llab/d\mul\,(r)$. 
In the ultra-relativistic transfer problem, practically all photons move 
forward in radius and cross a given $r$ only once.
We accumulate the statistics of angles and energies of photons at radius $r$, 
which gives $d\Llab/d\mul$ and the intensity of radiation in the lab frame,
$\Ilab=(4\pi r^2\mul)^{-1}d\Llab/d\mul$. The corresponding intensity in the 
fluid frame is given by $\I=\D^{-4}\Ilab$ where $\D$ is the Doppler factor 
(eq.~\ref{eq:nu}). Open circles in Figure~1 show the result of this 
calculation at $r=\Rph/2$. It is in perfect agreement with the solution of 
the transfer equation. Similar excellent agreement is found at other radii.

The strong anisotropy of radiation at subphotospheric radii is a 
result of cooperating effects. Photons with large $\muc$ have 
a larger free path in the lab frame $\lambda$ (in the opaque zone, 
$\lambda\approx [1+\beta\muc]\,r/\tau$.) 
 The free path is accompanied by 
a shift in $\muc$, $\Delta\muc=(d\muc/ds)\,\lambda >0$. This leads to the 
pile up of photons along the radial direction. Photons with large $\muc$ 
also experience less adiabatic cooling than the more frequently scattered 
photons with small $\muc$.

\subsection{Fuzzy photosphere}

$\Rph$ was defined in equation~(\ref{eq:Rph}) as the radius where the 
parameter $\tau$ appearing in the transfer equation equals unity. 
It gives an estimate for the characteristic photospheric radius. 
Clearly, the sphere of radius $\Rph$ cannot be thought of as the 
last-scattering surface, for two reasons: (1) the optical depth seen by 
a photon depends on its emission angle $\muc=\cos\thc$ and (2) the free 
path of a photon near $\Rph$ is a random variable comparable to $\Rph$.
Therefore, the radius of last scattering $\rls$ is a random variable.
Its average value logarithmically diverges for a steady outflow extending 
to infinity and cannot be used to define the photosphere. 

The process of radiation decoupling from the scattering plasma may be described
by the probability distribution $dP/d\rls d\mucls$ where $\mucls=\cos\thcls$ 
is the emission angle of the photon (measured in the fluid frame) at the 
last-scattering point. Pe'er (2008) considered a similar distribution for 
$\rls$ and $\thlls$, where $\thlls$ is the emission angle in the lab frame. 
His analytical expression is however inaccurate. The correct expression is 
given in Appendix~B. Integrating $dP/d\rls d\mucls$ over $\mucls$, one finds 
the distribution of emitted photons over the radius of last scattering, 
$dP/d\rls$ (Fig.~4). 

The same $dP/d\rls$ is obtained using Monte-Carlo technique.
The advantage of the Monte-Carlo simulation is that it is easily extended
to hot outflows and to scattering with exact Compton cross section.
We calculated $dP/d\ln\rls$ for the collisionally heated jet in the model 
of B10. The result was practically identical to that shown in Figure~4.

%%%%%%%%%%%%%%%%%%%%%%%%%%%%%%%%%%%%%%%%%%
\begin{figure}
% \label{fig:I0}
% \begin{center}
% \epsscale{1.2}
\epsscale{1.135}
\plotone{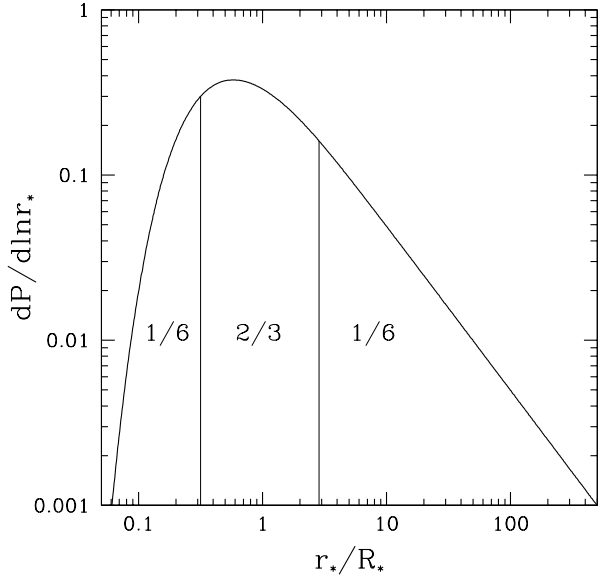}
\caption{Distribution of the last-scattering radius $\rls$, obtained by 
integrating equation~(\ref{eq:dist1}) over $\mucls$. 
Identical distribution is obtained using the Monte-Carlo technique.
The distribution remains practically the same in all models of photon 
transfer calculated in this paper: isotropic scattering, Thomson scattering in 
cold plasma, and Compton scattering in hot plasma. Approximately 2/3 of 
photons have $\rls$ between $0.3\Rph$ and $3\Rph$.
}
% \end{center}
\end{figure}
%%%%%%%%%%%%%%%%%%%%%%%%%%%%%%%%%%%%%%%%%%

In view of the broad distribution of $\rls$, it is not appropriate to 
locate the GRB photosphere at any specific radius, as emphasized 
by Pe'er (2008). When a characteristic radius is needed for rough estimates, 
$\Rph$ would be a reasonable choice. Alternatively, a characteristic 
photosphere could be defined as the sphere outside of which 50\% of photons 
are released (i.e. experience the last scattering). The corresponding 
radius is $0.8\Rph$.

%#########################################################################

\section{Transfer with exact electron scattering}
\medskip

Even for cold outflows, the radiative transfer is not exactly described by 
the model of coherent isotropic scattering. Two effects contribute to this: 
(1) electron scattering is not isotropic, and (2) radiation becomes 
polarized in the process of radiative transfer, with two modes of 
polarization, so two equations describe the transfer problem instead of one. 

Furthermore, if the outflow is strongly heated (as expected in GRBs) 
scattering is not coherent, i.e. does not conserve photon energy in the 
fluid frame. This has a strong effect on the intensity of radiation, 
as instead of passive adiabatic cooling radiation is heated through the 
Comptonization process.

In \Sec~4.1 we develop the accurate transfer model for Thomson
scattering in a cold outflow, which takes into account polarization. 
In \Sec~4.2 we present a transfer model for heated outflows. The results 
are compared with the simple isotropic-scattering model of \Sec~3.

\subsection{Thomson scattering and polarization}

The polarized transfer in a static, cold electron medium was described
by Chandrasekhar (1960) and Sobolev (1963). In axisymmetric problems, 
e.g. in plane-pallel or spherical geometries, there are two polarization 
modes of radiation: one with electric field perpendicular to the plane
containing the photon direction and the axis of symmetry and the other 
with electric field parallel to this plane. Let $I_\perp$ and $I_\parallel$ 
be the energy intensities of the two modes. Scattering can change the 
polarization state, so four scattering processes can occur: 
$\perp\rightarrow \perp$, $\perp\rightarrow \parallel$, 
$\parallel\rightarrow \perp$, and $\parallel\rightarrow \parallel$.
They are described by four different cross-sections.
The total intensity of radiation is $\I=\I_\perp+\I_\parallel$.
The degree of polarization is $p=\Q/\I$ where $\Q=\I_\perp-\I_\parallel$. 

The transfer equations for $\I$ and $\Q$ in a static medium are given in 
Chandrasekhar (1960) and Sobolev (1963). The generalization of these 
equations to the case of a relativistically moving medium is straightforward 
(see Beloborodov 1998 for equations in the plane-parallel geometry).
Here we are interested in spherical ultra-relativistic outflows. 
We will use the corresponding equations for the intensities in fluid frame, 
which are well-behaved in the limit $\beta\rightarrow 1$.
The transfer equations for $\I$ and $\Q$ are similar to 
equation~(\ref{eq:trans}),\footnote{
    Note that the polarization states are invariant under Lorentz boosts along 
    the axis of symmetry. Therefore, $\I$ and $\Q$ 
    obey the same transformation between the fluid frame and lab frame:
    $\Ilab=\D^4\I$ and $\Qlab=\D^4\Q$, where $\D$ is the Doppler factor 
    (see Appendix~A). The source functions $\S$ and $\R$ transform in the 
    same way.}
\be
\label{eq:I_Th}
   \frac{\partial\I}{\partial \ln r}
  =-\left(1-\muc^2\right)\,\hh\,\frac{\partial\I}{\partial\muc}
      - 4\,(1-\muc\hh)\,\I +\tauT\,\frac{(\S-\I)}{1+\muc},
\ee
\be
\label{eq:Q_Th}
   \frac{\partial\Q}{\partial \ln r}
  =-\left(1-\muc^2\right)\,\hh\,\frac{\partial\Q}{\partial\muc}
      - 4\,(1-\muc\hh)\,\Q +\tauT\,\frac{(\R-\Q)}{1+\muc}.
\ee
Here $\S$ and $\Q$ are the source functions in the fluid frame, where the 
scattering medium is static. Their expression in terms of moments of $\I$ 
and $\Q$ is exactly the same as in the static problem described by 
Chandrasekhar and Sobolev,
\be
\label{eq:S_Th}
   \S=\I_0+\frac{3}{8}\left(3\muc^2-1\right)
           \left(\I_2-\frac{\I_0}{3}+\Q_0-\Q_2\right),
\ee
\be
\label{eq:R_Th}
   \R=\frac{9}{8}\left(1-\muc^2\right)
           \left(\I_2-\frac{\I_0}{3}+\Q_0-\Q_2\right).
\ee
The parameter $\tauT$ appearing in equations~(\ref{eq:I_Th}) and 
(\ref{eq:Q_Th}) is defined using Thomson cross section $\sT$,
\be
   \tauT(r)\equiv \frac{\sT\dN_e}{4\pi r\,\g^2\,\beta c}, \qquad
   \Rph=\frac{\sT\,\dN_e}{4\pi\,\g^2\,\beta c},
\ee
where $r=\Rph$ corresponds to $\tauT=1$. 

%%%%%%%%%%%%%%%%%%%%%%%%%%%%%%%%%%%%%%%%%%
\begin{figure}
% \label{fig:I0}
% \begin{center}
% \epsscale{1.2}
\epsscale{1.135}
\plotone{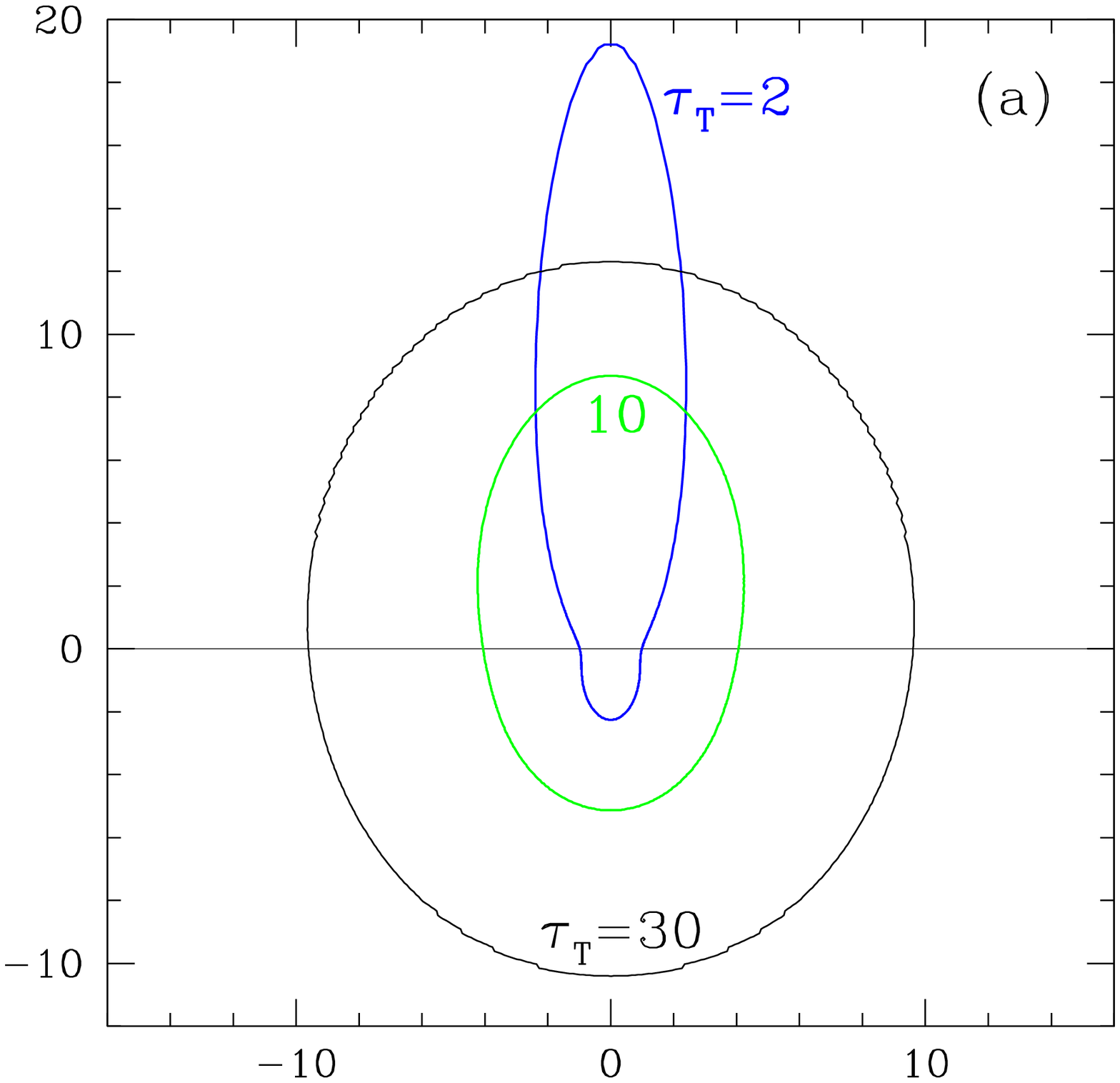}
\plotone{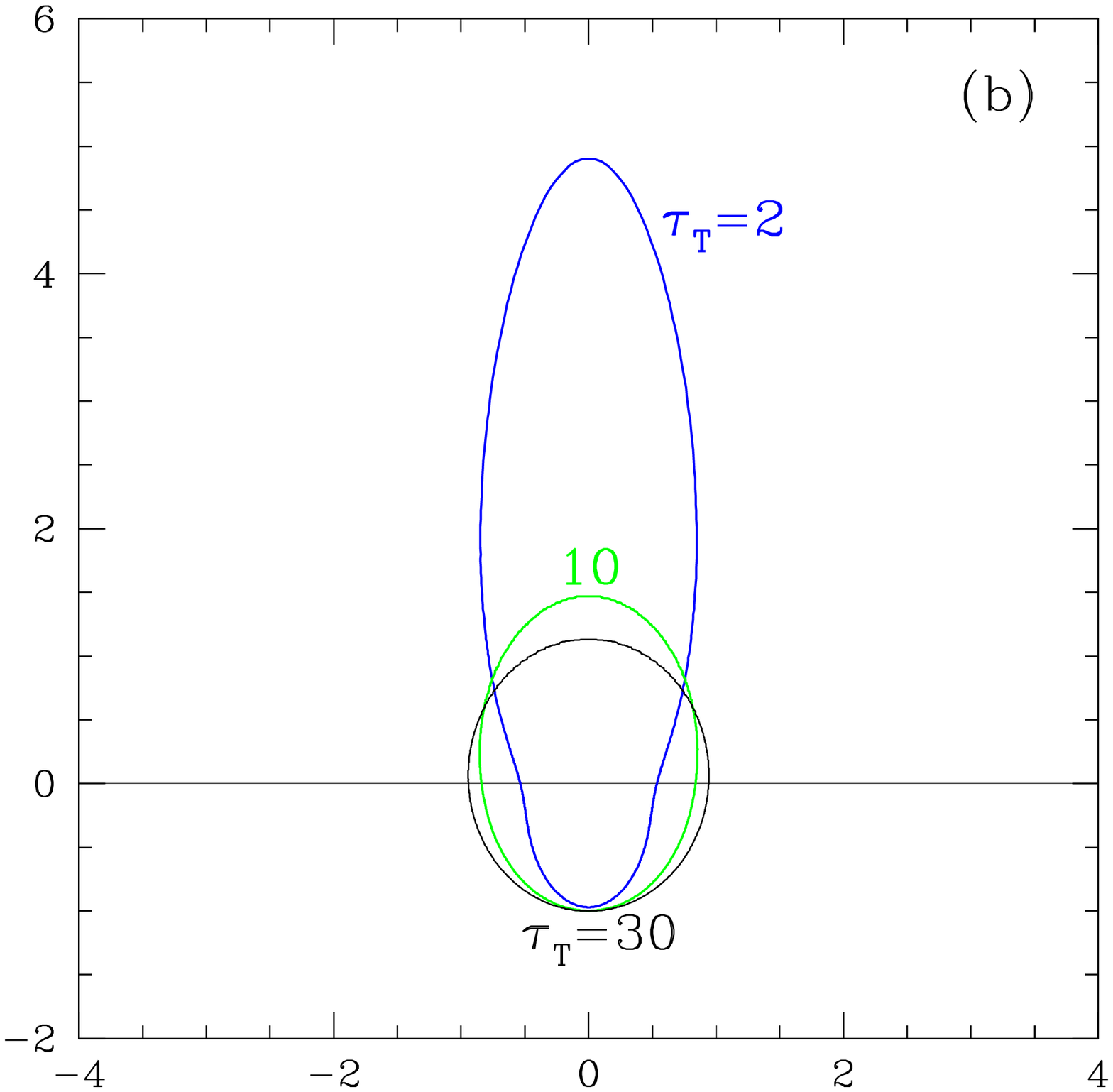}
\caption{
(a) Same as Fig.~1 but for Thomson scattering and taking into account 
polarization. The plot shows intensity $\I(\muc)$ (in the fluid frame)
at three radii: $r/\Rph\approx 0.03$, $0.1$ and $0.5$.
(b) The corresponding solution for the photon-number 
intensity $\IN(\muc)$. It is normalized so that $\IN_1+\beta\IN_0=1$ 
(note that $\IN_1[r]+\beta\IN_0[r]=const$ expresses conservation of
photon number, see \Sec~2.3); $\beta=1$ in our transfer problem.
}
% \end{center}
\end{figure}
%%%%%%%%%%%%%%%%%%%%%%%%%%%%%%%%%%%%%%%%%%

%%%%%%%%%%%%%%%%%%%%%%%%%%%%%%%%%%%%%%%%%%
\begin{figure}
% \label{fig:I0}
% \begin{center}
% \epsscale{1.2}
\epsscale{1.135}
\plotone{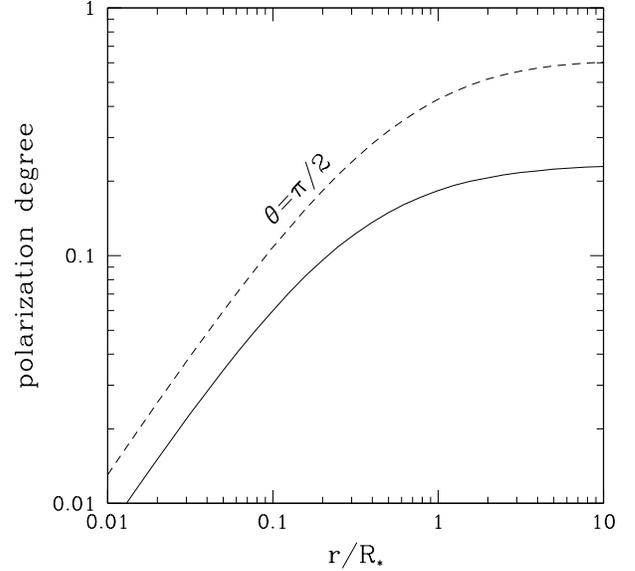}
\caption{
Polarization degree of radiation transferred through the outflow with 
Thomson scattering opacity, $p=Q/I$. Dashed curve shows the polarization of 
radiation propagating at angle $\thc=\pi/2$ in the fluid frame. Solid curve 
shows the angle-averaged polarization defined in equation~(\ref{eq:pav}).
}
% \end{center}
\end{figure}
%%%%%%%%%%%%%%%%%%%%%%%%%%%%%%%%%%%%%%%%%%

Similar to \Sec~3, we will consider here outflows with $\dN_e(r)=const$ 
and $\g(r)=const$. This implies $\tauT(r)=\Rph/r$ and $\hh=1$.
Equations~(\ref{eq:I_Th}) and (\ref{eq:Q_Th}) are solved numerically 
in the same way as described in \Sec~3.1.
The resulting angular distribution of intensity $\I$ in the fluid frame 
is shown in Figure~5a. It is similar to the model with isotropic scattering 
(Fig.~1). 

The polarization degree of radiation is shown in Figure~6. It begins to 
grow in the subphotospheric region (where anisotropy develops), and 
radiation becomes significantly polarized in the photospheric region. 
The strongest polarization $p\approx 0.6$ is found at large radii for 
radiation at angles $\theta\approx \pi/2$.  It is produced by scattering 
of strongly beamed radiation, which naturally generates a high 
polarization at scattering angles close to 90~degrees. The overall 
polarization at a given radius may be described by the average $p$
that is defined using the energy fluxes in the two modes measured in the 
lab frame. This definition involves the transformation of moments $\I_m$
and $\Q_m$ to the lab frame, which gives
\be
\label{eq:pav}
  <p>=\frac{\Qlab_1}{\Ilab_1}=\frac{\beta(\Q_0+\Q_2)+(1+\beta^2)\Q_1}
                                   {\beta(\I_0+\I_2)+(1+\beta^2)\I_1}.
\ee
This expression is well-behaved in the limit $\beta\rightarrow 1$ and 
becomes $<p>=(\Q_0+2\Q_1+\Q_2)/(\I_0+2\I_1+\I_2)$.
It reaches 0.24 outside the photosphere (Fig.~6).

Equations~(\ref{eq:I_Th}) and (\ref{eq:Q_Th}) govern the transfer of 
{\it energy} in the two polarization modes. The photon {\it number} in 
the two modes is described by intensities $\IN_\perp$ and $\IN_\parallel$. 
The equations for $\IN=\IN_\perp+\IN_\parallel$ and 
$\QN=\IN_\perp-\IN_\parallel$ read (cf. the similar eq.~\ref{eq:trans_num}) 
\be
\label{eq:IN_Th}
   \frac{\partial\IN}{\partial \ln r}
  =-\left(1-\muc^2\right)\,\hh\,\frac{\partial\IN}{\partial\muc}
      - 3\,(1-\muc\hh)\,\IN +\tauT\,\frac{(\SN-\IN)}{1+\muc},
\ee
\be
\label{eq:QN_Th}
   \frac{\partial\QN}{\partial \ln r}
  =-\left(1-\muc^2\right)\,\hh\,\frac{\partial\QN}{\partial\muc}
      - 3\,(1-\muc\hh)\,\QN +\tauT\,\frac{(\RN-\QN)}{1+\muc}.
\ee
The source functions $\SN$ and $\RN$ are related to the moments $\IN_m$
and $\QN_m$ ($m=0,1,2$) in the same way as $\S$ and $\R$ are related to 
$\I_m$ and $\Q_m$ (eqs.~\ref{eq:S_Th} and \ref{eq:R_Th}).

We numerically solved equations~(\ref{eq:IN_Th}) and (\ref{eq:QN_Th}) 
with $\hh=1$. The resulting angular distribution $\IN(\muc,r)$ is shown 
in Figure~5b. It is less anisotropic than the energy intensity $\I(\muc,r)$. 
(A similar result is found in the isotropic-scattering model.) 
Correspondingly, the quantity $\QN/\IN$ is smaller than $p=\Q/\I$, 
roughly by a factor of $\sim 3/4$.

%%%%%%%%%%%%%%%%%%%%%%%%%%%%%%%%%%%%%%%%%%
\begin{figure}
% \label{fig:I0}
% \begin{center}
% \epsscale{1.2}
\epsscale{1.135}
\plotone{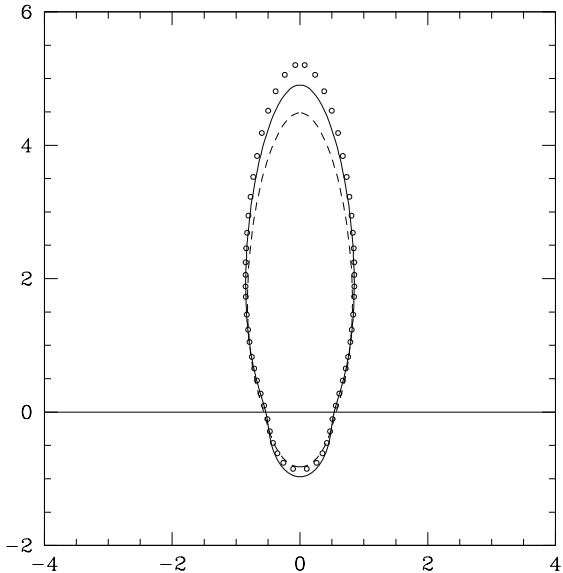}
\caption{
Comparison of photon-number intensity $\IN(\muc)$ at $r=\Rph/2$ 
in three models: isotropic coherent scattering (dashed curve),
polarized Thomson scattering in a cold outflow (solid curve), and 
unpolarized Compton scattering in a hot outflow (open circles), see
text for the description of the hot-outflow model.
In all three models, the intensity is normalized so that the conserved
quantity $\IN_1+\beta\IN_0$ equals unity. The models assume $\beta\approx1$.
}
% \end{center}
\end{figure}
%%%%%%%%%%%%%%%%%%%%%%%%%%%%%%%%%%%%%%%%%%

\subsection{Scattering in a hot plasma}

GRB jets are heated, which leads to Comptonization --- a significant flow 
of heat from particles to radiation. This process depends on the electron 
temperature that must be calculated self-consistently. For example, 
consider the fiducial model of the collisionally-heated jet in B10.
The jet with Lorentz factor $\g=600$ is heated at radii $r>R_n\approx\Rph/20$, 
and its electron temperature outside $R_n$ may be approximated by
\be
\label{eq:Theta_e}
  \Theta_e\equiv\frac{kT_e}{m_ec^2}=0.045\,\left(\frac{r}{\Rph}\right)^{0.23}.
\ee
The initial temperature of radiation at $R_n$ is 0.6~keV. Comptonization 
begins at $R_n$ with Kompaneets' $y$-parameter $y=4\Theta_e\tauT\sim 1$. 
Equation~(\ref{eq:Theta_e}) is a reasonable approximation in the main heating 
region $\Rph/20 < r <\Rph$. The exact value of temperature outside $\Rph$ is 
not important --- its contribution to thermal Comptonization is small, --- 
and the above approximation for $T_e$ will be sufficient for our purposes.

Comptonization significantly changes the energy intensity $\I(\muc,r)$ 
compared with the cold-jet model. 
In addition to the thermal plasma, nonthermal particles are continually 
injected with energies $\sim 140$~MeV from the decay of pions produced by 
nuclear collisions. These particles convert their energy to a small number 
of high-energy photons, which impact the radiative transfer and the observed 
spectrum above 20~MeV (B10).

Here, however, we limit our consideration to the 
transfer of photon {\it number} (rather than energy). 
The high-energy photons make a negligible contribution to the number 
intensity $\IN(\muc,r)$, so it is sufficient to consider scattering 
by the thermal plasma with temperature (\ref{eq:Theta_e}), which strongly 
dominates the optical depth. The solution for the 
number intensity $\IN(\muc,r)$ turns out to be close to that in a cold jet. 
The obtained angular distribution of Comptonized photons at $r=\Rph/2$ is 
shown by open circles in Figure~7. Its beaming is somewhat stronger 
compared with the cold-jet model of \Sec~4.1, however the difference is 
modest. The effect of electron heating on the transfer solution for photon 
number is $\sim 10$\%.

%########################################################################

\section{Radiation-dominated outflow}
\medskip

Anisotropic radiation always tends to push the flow toward the equilibrium 
velocity at which the net flux of radiation vanishes in the fluid frame. 
When this effect is strong, it leads to the peculiar ``equilibrium transfer''
where radiation and plasma self-organize to flow with a common velocity
(Beloborodov 1998, 1999). In this section, we discuss the conditions for 
this regime and the corresponding solution of the transfer problem.

The radiative force applied to each electron (or positron) in the fluid 
frame is given by $f=4\pi\I_1\sT/c$. We assume that the electron thermal 
motion in the fluid frame is slow (non-relativistic).\footnote{ 
      GRB outflows start with a relativistic temperature $kT\sim 1$~MeV, 
      but they are quickly cooled by adiabatic expansion to much lower 
      temperatures before they reach the photospheric radius. Compton 
      scattering provides a strong thermal coupling between the plasma and 
      radiation and keeps the electron temperature relatively low 
      (non-relativistic).}
Then Lorentz transformation gives the same value for the force measured in 
the lab frame, $\tilde{f}=f$. The outflow acceleration is governed by the 
dynamic equation,
\be
\label{eq:acc}
  \frac{\rho}{n}\,c^2\beta\,\frac{d(\g\beta)}{dr}=4\pi\I_1\,\frac{\sT}{c}.
\ee
Here $\I_1$ is the first moment of radiation in the fluid frame 
(eq.~\ref{eq:mom}), $\rho$ is the rest-mass density of the plasma, and $n$ is 
the number density of $e^\pm$; both $\rho$ and $n$ are measured in the fluid 
frame. Then, for ultra-relativistic outflows ($\beta\rightarrow 1$), one finds
\be
\label{eq:h1}
   \hh=1-\frac{d\ln\g}{d\ln r}=1-\frac{4\pi\I_1}{\rho c^3}\,\tauT.
\ee
One can use the approximation $\hh=1$ (i.e. $\g=const$) if the last term in 
equation~(\ref{eq:h1}) is much smaller than unity. At the photospheric radius 
$\Rph$ this term equals $4\pi\I_1\,(\rho c^3)^{-1}=\chi\,(U/\rho c^2)$ 
where $\chi=\I_1/\I_0<1$ is a numerical factor that depends on the angular 
distribution of radiation and $U=4\pi\I_0/c$ is the radiation energy density 
in the fluid frame. If $U/\rho c^2$ is not small compared with unity, 
the transfer equation should be solved with the self-consistent function 
$\hh(r)$ given by equation~(\ref{eq:h1}). 

The self-consistent solution of the transfer problem can be obtained 
analytically in the extreme radiation-dominated regime $U/\rho c^2\gg 1$. 
Note that there exists a special value of $\g=\gsat$ for which the radiative 
force (eq.~\ref{eq:acc}) vanishes; this value corresponds to $\I_1=0$. 
The force is positive if $\g<\gsat$ and negative if $\g>\gsat$, i.e. it 
always pushes the outflow toward $\g=\gsat$. In the radiation-dominated 
regime, the timescale for dynamical relaxation toward $\g=\gsat$
is shorter than the outflow expansion timescale. This means that the
outflow maintains $\g\approx \gsat$ and $\I_1\ll\I_0$.

In this regime, the ultra-relativistic transfer has a simple solution,
\be
\label{eq:sol}
   \I(\muc,r)=\frac{C}{r^4}, \qquad \Q(\muc,r)=0, \qquad \hh=0,
\ee
where $C$ is a constant determined by the inner boundary condition.
It is straightforward to verify that equation~(\ref{eq:sol}) is the 
solution of equations~(\ref{eq:I_Th}) and (\ref{eq:Q_Th}). The first 
moment $\I_1$ is a small (next-order) quantity. It controls the outflow 
acceleration $\g\propto r$ and is given by 
$\I_1=(4\pi\tauT)^{-1}\rho c^3\ll \I_0$.
The outflow accelerates linearly with radius both inside
and outside the photosphere, as long as $U\gg\rho c^2$. Note that
$U/\rho c^2\propto r^{-1}$, so the outflow eventually must reach radii 
where $U<\rho c^2$ and the acceleration ends. This transition occurs 
outside $\Rph$ and has no effect on radiation escaping the outflow to 
distant observers.

Equation~(\ref{eq:sol}) states that radiation remains isotropic 
in the fluid frame which accelerates as $\g\propto r$.
The sustained isotropy is a consequence of a remarkable fact: 
a freely propagating photon between two successive scatterings at radii 
$r_1$ and $r_2$ does not change its angle measured in the fluid frame,
$\thc_1=\thc_2$. This fact can be derived as follows. In the lab frame 
the angle of a photon propagating from $r_1$ to $r_2$  satisfies the 
equation of a straight line, $r_1\sin\thl_1=r_2\sin\thl_2$.
Using $r_2/r_1=\g_2/\g_1$ and Doppler transformation $\sin\thc=\D\sin\thl$
(where $\D=\g[1+\cos\thc]$ when $\beta\rightarrow 1$), 
one finds $\thc_1=\thc_2$.

Thus, free propagation between scatterings does not generate
any change in the angular distribution of photons in the fluid frame,
and an initially isotropic radiation remains isotropic, $\I=\I_0$.
Scattering of isotropic radiation gives isotropic radiation, so the 
source function also remains isotropic in the fluid frame, $\S=\I_0$.
Naturally, the isotropic radiation remains unpolarized. 

It is easy to see why the energy intensity in the fluid frame scales with 
radius as $\I\propto r^{-4}$. First note that conservation of photon number 
implies that the photon-number intensity scales as 
$\IN\propto \g^{-1}r^{-2}\propto r^{-3}$ (see \Sec~2.3 and use isotropy, 
$\IN\approx\IN_0$ and $\IN_1\ll \IN_0$). Coherent scattering does not 
affect the photon energy in the fluid frame, so $\nu$ changes only 
during free propagation of the photon.
Propagation between successive scatterings at $r_1$ and $r_2$ occurs
with constant energy in the lab frame, $\D_2\,h\nu_2=\D_1\,h\nu_1$. 
Using $\thc_2=\thc_1$, one finds $\nu_2/\nu_1=\D_2/\D_1=\g_2/\g_1=r_2/r_1$. 
Thus, $\nu\propto r^{-1}$ for each photon.
Together with $\IN\propto r^{-3}$ this implies $\I\propto r^{-4}$.

The existence of a frame (fluid frame) where radiation remains isotropic and 
scattering is coherent implies that scattering has no effect on radiation, 
i.e. the transfer occurs as if radiation propagated in vacuum. Indeed, 
consider the basic transfer equation~(\ref{eq:transfer}) in the lab frame. 
Coherent isotropic scattering in the fluid frame gives $\Sc=\Ic$, which
implies $\Sl=\Il$ in the lab frame and hence $d\Il/ds=0$. Thus, 
radiation intensity in the lab frame remains constant along the ray, 
just like propagation in vacuum. In particular, the spectrum of radiation 
is preserved and its beaming angle decreases as $r^{-1}$.

The corresponding solution for the intensity in the fluid frame 
can be obtained by the Doppler transformation $\Ic=\D^{-3}\Il$ (Appendix~A)
of the vacuum solution for $\Il$. Alternatively, the same result can be 
obtained from equation~(\ref{eq:transfer_c}). Substituting $\hh=0$, one gets
\be
  \Ic(\nuc,\muc,r)=\Ic\left(\frac{\nuc\,r}{\rin},\muc,\rin\right)\,
                   \left(\frac{r}{\rin}\right)^{-3}.
\ee
It confirms that, when viewed in the fluid frame, the transfer preserves 
isotropy of photon distribution and shifts each photon in frequency as 
$r^{-1}$. 
 
To summarize, as long as $U\gg \rho c^2$ radiation behaves as if there 
were no scattering and it streamed freely, regardless of the optical depth. 
This is a special feature of the ultra-relativistic transfer in spherical 
geometry. It differs from the radiation-dominated transfer with $\g=\g_0$ 
in the plane-parallel geometry (Beloborodov 1998; 1999).

%########################################################################

\medskip

\section{Variable jets and the steady spherically symmetric model}

In this paper, we considered radiative transfer in outflows that are 
steady and spherically symmetric. Here we discuss why these assumptions 
are not so restrictive as they might seem and the model may describe 
variable jets. 

Radial outflows with Lorentz factors $\g\gg 1$ have two well-known features:
(1) Their parts are causally disconnected on scales larger than 
$l_\perp\sim r/\g$ on any sphere of radius $r$ ($r\sim\Rph$ should be taken 
as the characteristic radius for the problem of photospheric emission). 
(2) Since both radiation and fluid move outward with almost speed of light, 
the radial diffusion of radiation relative to the fluid is inefficient 
on scales $\delta r\gg l_\parallel$ where $l_\parallel\sim r/\g^2$. 
To a first approximation, each ``elementary pancake'' of volume 
$l_\parallel\times l_\perp\times l_\perp$ in the lab frame has its own 
radiative transfer and produces photospheric emission almost independently 
from the neighboring pancakes.
If strong inhomogeneities of the jet are confined to scales much larger 
than $l_\parallel$ and $l_\perp$ (in the radial and transverse directions, 
respectively), radiative transfer in each pancake occurs as if it were 
part of a steady, spherically symmetric outflow. 

The independence of emissions from different pancakes can be better 
quantified if one considers the photon exchange between two pancakes 
separated by $\delta r>l_\parallel$. The exchange is one-way only: the 
trailing pancake can receive photons from the leading pancake
(the opposite communication is impossible for $\delta r>l_\parallel$).
This ``trailing diffusion'' of radiation was studied by Pe'er (2008).
He considered a very narrow shell of photons (formally a delta-function 
of radius) injected at a small $r$ in a steady jet with a constant Lorentz 
factor. The photons diffuse through the jet and eventually escape, producing 
an isolated pulse of emission that will be received by a distant observer.
The characteristic observed width of this pulse is 
$\delta t_{\rm obs}\sim \tph=l_\parallel/c\sim \Rph/\g^2c$, and it 
has an extended tail whose intensity decreases as $(\tobs/\tph)^{-2}$. 
This implies that the trailing diffusion of radiation on scales
$\delta r>l_\parallel$ is suppressed as $(\delta r/l_\parallel)^{-2}$.

A realistic GRB jet is continually filled with thermal radiation near the 
central engine. It may be viewed as a continual sequence of elementary 
pancakes that release their photons near $\Rph$. The strong Doppler beaming 
implies that photospheric emission seen by a distant observer is dominated 
by a small patch $l_\perp\times l_\perp$ on the sphere of radius $\sim\Rph$.
The observer receives radiation released by consecutive pancakes in the 
same order as they pass through $\Rph$. The observed timescale of passage 
of one elementary pancake through $\Rph$ is $\tph$, which may be smaller
than 1~ms for GRBs. The steady transfer model developed in this paper is 
valid for GRBs with variability timescales $\delta t_{\rm obs}>\tph $.
The model permits different $\Rph$ for pancakes separated by timescales 
$\delta t_{\rm obs}=\delta r/c\gg\tph$.

%##########################################################################

\medskip

\section{Discussion}
\medskip

This paper explored radiative transfer in ultra-relativistic outflows. 
The transfer problem is well defined and simplifies in the limit $\g\gg 1$. 
In this limit, radiation propagating backward in the lab frame can be 
neglected. Therefore, the transfer solution is independent of the outer 
boundary condition, in contrast to transfer in static media studied by 
Chandrasekhar (1960) and Sobolev (1963). The problem is solved by direct 
integration of the transfer equation, with no need for iterations. 
The model with $\g\rightarrow\infty$ gives excellent approximation 
to transfer in outflows with finite $\g>10$.

\subsection{Transfer in radiation-dominated and matter-dominated outflows}

The approach developed in this paper gives a simple solution for the old 
problem of radiation-dominated jet discussed by Paczy\'nski (1986) and 
Goodman (1986). This jet is baryon-clean. It is very opaque at small radii 
because of the thermal population of $e^\pm$ pairs. Almost all pairs 
annihilate at larger radii, and almost all the jet energy is carried by 
radiation that is released at the photosphere. Paczy\'nski and Goodman 
considered the opaque zone of the radiation-dominated jet and derived its 
Lorentz factor $\g\propto r$ from energy-momentum conservation. They argued 
that quasi-thermal emission should be observed from the jet, with a spectral 
peak near 1~MeV. They suggested, however, that the observed spectrum 
should be different from blackbody because of complicated transfer effects 
near the photosphere. Goodman (1986) performed a numerical calculation 
with simplifying assumptions, which gave a nonblackbody spectrum.

In fact, the exact transfer solution for this problem gives precisely 
blackbody spectrum. As shown in \Sec~5, photons in a radiation-dominated
jet are transferred as if there were no scattering at all.
The radiation remains isotropic in the fluid frame which accelerates
as $\g\propto r$ both inside and outside the photosphere.
A distant observer can think that radiation freely propagates from the 
central engine of the jet.\footnote{The only deviation from the 
    free-propagation solution occurs where the jet temperature drops below 
    $\sim m_ec^2$ and the equilibrium density of pairs drops below the 
    density of photons. In this region, radiation receives significant 
    energy from the annihilated pairs, which boosts its density by the 
    factor of 11/4 (similar to what happens in the expanding universe). 
    After this transition the jet is still extremely opaque due to the 
    remaining (exponentially reduced) $e^\pm$ population, and the radiation 
    remains Planckian.}
Note also that regardless of how strong dissipation/heating may occur in 
the jet it does not have a dramatic impact on the shape of the spectral
peak, because the energy budget of heating is negligible compared with 
the Planck radiation. The observed radiation should have a blackbody 
spectrum. 

The opposite, ``matter-dominated'' regime was considered by Paczy\'nski 
(1990). In the opaque zone radiation cools adiabatically
and the jet energy becomes dominated by baryons. Then it continues inertial 
expansion (coasting) with some relict thermal radiation in it until the 
radiation is released at the photosphere. 

In the matter-dominated regime, the photospheric spectrum cannot have the 
blackbody shape, even if the outflow cools passively, with no heating, 
up to the photosphere. B10 showed that the photospheric spectrum in the 
soft X-ray band has the slope $\alpha\approx 0.4$ instead of the blackbody 
(Rayleigh-Jeans) slope $\alpha=1$. Moreover, collisional heating in GRB
jets transforms the photospheric spectrum into the Band-type radiation,
with extended high-energy emission instead of the exponential cutoff
above 1~MeV. Thus, the photospheric spectrum of a matter-dominated jet 
is changed from blackbody both below and above the MeV peak. 

Besides giving a non-blackbody spectrum, the radiative transfer in 
matter-dominated jets has other interesting features. Radiation becomes 
strongly anisotropic in the fluid frame well before it decouples from 
the fluid. Radiation at the characteristic photospheric radius $\Rph$ has 
the beaming factor $b\sim 30$ (Fig.~2). Beaming affects the adiabatic cooling 
of photons in the subphotospheric region. The net cooling factor for 
radiation emitted at a radius $\rin\ll\Rph$ equals $2(\rin/\Rph)^{2/3}$.

In a heated jet, adiabatic cooling of radiation is counter-balanced (or 
dominated) by Comptonization, so the mean photon energy can grow with radius. 
This has a strong effect on the transfer solution for the radiation intensity.
However, if one focuses on the transfer of photon number (rather than energy), 
the results are not sensitive to heating. The photon-number intensity 
$\IN$ in the passively cooling and heated jets is very similar 
(the difference is $\sim 10$\%). In both cases, $\IN$ is strongly beamed 
in the subphotospheric region (Fig.~7).

\subsection{Detecting the blackbody component in GRBs}

As discussed in \Sec~7.1, photospheric emission in GRBs can have a 
blackbody spectrum only when the photosphere is dominated by radiation 
(i.e. $U\gg\rho c^2$ at $r\sim\Rph$).
The detection of a blackbody component in a GRB spectrum would provide
clear evidence that part of the photospheric emission is in the 
radiation-dominated regime.

The existing data are inconclusive.
It includes GRB~090902B that was much discussed recently as a burst 
with a blackbody component. In fact, it is equally well fitted 
by the Band function plus a power law (Ryde et al. 2010).  
The data interpretation is further complicated by the variability of
photospheric emission. It may vary on very short timescales, as short as 
$\tph\sim \Rph/c\g^2$, which can be smaller than one millisecond for a 
typical GRB. The achieved temporal resolution of 
spectral analysis is far worse than 1~ms and may not give the true 
instantaneous photospheric spectrum. The photospheric emission may
quickly switch between the radiation-dominated and matter-dominated 
regimes and these variations would remain undetected.

The low-energy slopes of the observed GRB spectra, $\alpha$, are 
affected by the time averaging, which tends to reduce $\alpha$. 
Bursts with largest $\alpha$ are most promising for detecting 
the blackbody component. In some cases, $\alpha\sim 1$ were reported 
% in GRB~911118, GRB~910807, GRB~910927, GRB~980306, and GRB~970111 
(Ghirlanda, Celotti \& Ghisellini 2003). This indicates the existence of 
the radiation-dominated regime.

\subsection{Detecting photospheric polarization}

Radiation remains unpolarized in radiation-dominated jets (\Sec~5).
In contrast, in matter-dominated jets, radiation acquires a strong 
linear polarization in the photospheric region (\Sec~4.1).
An ideal detector that has enough angular resolution to image the 
spherically-symmetric outflow on scales $\sim \g^{-1}\Rph$ would detect 
the polarization. In practice, such a high angular resolution is not 
achieved. The detectors receive a mixture of radiation whose 
polarization averages to zero unless something breaks spherical symmetry. 
 
Three principle possibilities for breaking the symmetry are as follows:
(1) The main emitting region of size $l_\perp\sim \g^{-1}\Rph$ is 
partially eclipsed.
(2) The outflow deviates from spherical symmetry on scales $\sim l_\perp$.
(3) The jet carries magnetic fields with a coherence scale $\simgt l_\perp$.
In magnetized jets, the synchrotron component of photospheric emission 
becomes dominant at photon energies below $\sim 100$~keV (B10; Vurm et al. 
2011).
This component can be highly polarized. Future polarization measurements 
across the X-ray spectrum will help estimate the magnetization of GRB jets.

\subsection{Modeling frequency-dependent radiative transfer in heated jets}

The photosphere of a baryonic jet is a fuzzy object --- about 2/3 
of photons are released in the region $\Rph/3<r<3\Rph$, and
the remaining 1/3 comes from even more extended region.
Modeling the heated anisotropic radiation emerging from this region 
requires accurate transfer simulations.

B10 developed a Monte-Carlo transfer code that solves the transfer 
problem in a broad range of photon energies up to 100~GeV, including 
the effects of $\gamma$-$\gamma$ absorption. Alternatively, one can 
use the kinetic method that solves the kinetic equations 
for the photon and electron distribution functions
 (Pe'er \& Waxman 2005; Vurm \& Poutanen 2009).
The developed kinetic codes have, however, one drawback: they assume 
isotropic radiation in the fluid frame, which is not a good approximation. 
Besides, it violates conservation of photon number in the lab frame. The 
kinetic method can be used more efficiently if it calculates the evolution 
of radiation by solving the transfer equation~(\ref{eq:transfer_c}).
This method will be implemented in an upcoming paper (Vurm et al. 2011). 

\acknowledgments
This work was supported by NSF grant AST-1008334 and NASA grant
NNX10AO58G.

%########################################################################

\appendix

\section{A. Basic equations of relativistic transfer}

Consider radiation with specific intensity $\Il(\nul,\mul,r)$ in the fixed
lab frame. Here $\nul$ is the photon frequency, $\mul=\cos\thl$, and $\thl$ 
is the photon angle with respect to the radial direction. Hereafter 
quantities measured in the fixed lab frame are denoted with tilde. 
The transfer equation reads
\be
\label{eq:transfer}
   \frac{d\Il}{ds}=\kl(\Sl-\Il),
\ee
where $ds$ is the path element along the ray and $\kl(\nul,\mul,r)$ is the 
absorption coefficient of the outflow in the lab frame. The source function 
$\Sl$ equals $\jl/\kl$, the ratio of emission and absorption coefficients 
(e.g. Chandrasekhar 1960).

The outflow in our problem is moving radially with velocity $\beta(r)$ and 
Lorentz factor $\g(r)$. The transfer equation in the lab frame is not 
well behaved in the ultra-relativistic limit $\g\gg 1$.
Therefore, we rewrite it in terms of intensity $\Ic$ measured in the 
fluid frame, i.e. in the frame comoving with the outflow.
This is straightforward to do using the usual transformation laws 
(Prokof'ev 1962; Castor 1972; Mihalas 1980). The transformations are given by 
\be
\label{eq:I}
   \Il=\D^3\Ic, \qquad \Sl=\D^3\Sc,
\ee
\be
\label{eq:nu}
   \nul=\D\nuc,   \qquad \D=\g(1-\beta\mul)^{-1}=\g(1+\beta\muc),
\ee
\be
\label{eq:mu}
   \mul=\frac{\muc+\beta}{1+\beta\muc}.
\ee
\be
\label{eq:k}
   \kl=\D^{-1}\kc.
\ee

Using $\Il/\nul^3=\Ic/\nuc^3$ and $\Sl/\nul^3=\Sc/\nuc^3$, 
equation~(\ref{eq:transfer}) may be written as
\be
\label{eq:transfer1}
  \nuc^3\,\frac{d}{ds}\left(\frac{\Ic}{\nuc^3}\right)
   =\frac{\kc}{\D}\left(\Sc-\Ic\right).
\ee 
$\Ic$ is considered as a function of $r$, $\muc$, $\nuc$,
and the derivative along the ray is expanded as 
\be
 \frac{d\Ic}{d s}=\frac{dr}{ds}\,\frac{\partial\Ic}{\partial r}
  +\frac{d\muc}{ds}\,\frac{\partial\Ic}{\partial\muc}
  +\frac{d\ln\nuc}{ds}\,\frac{\partial\Ic}{\partial \ln\nuc}.
\ee
Here one can use $dr/ds=\mul$, $d\nul/ds=0$, and $d\mul/ds=(1-\mul^2)/r$ 
(a consequence of $r\sin\thl=const$, which is valid for any straight line). 
The corresponding derivatives of $\nuc$ and $\muc$ are obtained using the 
transformations~(\ref{eq:nu}) and (\ref{eq:mu}). This gives
\begin{eqnarray}
  \frac{dr}{ds} &=& \frac{\g}{\D}\,(\muc+\beta),  \\
  \frac{d\muc}{ds}
  &=& \frac{1-\muc^2}{r}\left[1-\frac{\g}{\D\beta}\,(\muc+\beta) 
                        \,\frac{d\ln\g}{d\ln r}\right],  \\
 \frac{d\ln\nuc}{ds} &=& -\frac{\g\beta}{\D}\,\frac{(1-\muc^2)}{r}
                  -\frac{\g\,\muc}{\D\beta}\,(\muc+\beta)\,\frac{d\ln\g}{dr},
\end{eqnarray}
where we used $d\beta=d\g/\beta\g^3$ and $(1-\mul^2)=\D^{-2}(1-\muc^2)$.
Then equation~(\ref{eq:transfer1}) becomes,
\be
   (\muc+\beta)\,\frac{\partial\Ic}{\partial \ln r}
  +\left(1-\muc^2\right)\left[1+\beta\muc-\frac{(\muc+\beta)}{\beta}
          \,\frac{d\ln\g}{d\ln r}\right]\,\frac{\partial\Ic}{\partial\muc}
      -\left[\beta\left(1-\muc^2\right)+\frac{\mu(\mu+\beta)}{\beta}
          \,\frac{d\ln\g}{d\ln r}\right]
      \left(\frac{\partial\Ic}{\partial\ln\nuc}-3\Ic\right)
  =\tau_\nu(\Sc-\Ic),
\label{eq:transfer_c1}
\ee
where
\be
\label{eq:tau_nu}
   \tau_\nu(r,\nuc)\equiv\frac{\kc(r,\nuc)\, r}{\g}.
\ee
Equation~(2.12) in Mihalas (1980) is reduced to 
equation~(\ref{eq:transfer_c1}) in the steady case.

%########################################################################

\section{B. Analytic solution}

In \Sec~2 we solved the transfer equation numerically. Here we collect useful 
analytical formulas that may be used instead of the numerical solution.

\subsection{Optical depth along the ray}

Consider a photon propagating from radius $r_1$ to $r_2$ along a straight
line in the lab frame. Let $\thl_1$ be photon angle at $r_1$.
The optical depth along the ray from $r_1$ to $r_2$ is 
\be
\label{eq:tau1}
  \taur(r_1,\thl_1\symb r_2)=\int_{r_1}^{r_2} \tilde{\kappa}(r,\thl)\,
                               \frac{dr}{\cos\thl}.
\ee
Here $\thl(r)$ is photon angle at radius $r$. It satisfies the relation 
(which expresses the fact the photon moves along a straight line),
\be
\label{eq:ray}
   r\sin\thl=r_1\sin\thl_1.
\ee
The scattering opacity in the lab frame is given by 
$\tilde{\kappa}=\D^{-1}\kappa=\g(1-\beta\cos\thl)\,\sigma n$.
Let us consider an outflow with $\g(r)=const$ and $n\propto r^{-2}$.
Then the elementary integral in equation~(\ref{eq:tau1}) gives
\be
\label{eq:tau2}
 \taur(r_1,\thl_1\symb r_2)=\tau(r_1)\,\g^2
               \left[\frac{\thl_1-\thl_2}{\sin\thl_1}
                     -\beta\left(1-\frac{r_1}{r_2}\right)\right],
\ee
where $\tau(r)\equiv n\sigma r/\g$.
If $\g\gg 1$, one can expand equation~(\ref{eq:tau2}) in $\g^{-1}$,
\be
\label{eq:tau3}
   \taur(r_1,\x_1\symb r_2)=\frac{\tau(r_1)}{6}
    \left(1-\frac{r_1}{r_2}\right)\left[3+\left(1+\frac{r_1}{r_2}
             +\frac{r_1^2}{r_2^2}\right)\,\x_1\right] 
             +{\cal O}\left(\g^{-2}\right).
\ee
Here $\x={\cal O}(1)$ is a convenient variable related to the photon angle,
\be
\label{eq:x}
   x\equiv\g^2\thl^2=\frac{1-\muc}{1+\muc},
\ee
where $\muc=\cos\thc$ is measured in the fluid frame.

If we take $r_2\rightarrow\infty$ and drop index ``1'' for the emission
point, equations~(\ref{eq:tau2}) and (\ref{eq:tau3}) are reduced to  
\be
\label{eq:tau4}
   \taur(r,x\symb \infty)=\tau(r)\,\g^2\left(\frac{\thl}{\sin\thl}
                                 -\beta\right)
        =\frac{\tau(r)}{6}\left(3+\x\right) + {\cal O}\left(\g^{-2}\right).
\ee
A similar formula for the optical depth along the ray from radius $r$ to 
infinity is given in Abramowicz et al. (1991) and Pe'er (2008).

\subsection{Expressions for intensity and source function}

The formal solution for the transfer problem is written in terms of the 
source function. Let us first consider the transfer of photon number
(\Sec~2.3). The corresponding intensity in the lab frame is given by 
\be
\label{eq:formal0}
   \INlab(\mul,r)=\int_0^\infty \SNlab(\mul_1,r_1)\,
   \exp\left[-\taur(r_1,\mul_1\symb r)\right]\,d\taur(r_1,\mul_1\symb r),
\ee
where $r_1<r$ is running along the ray and $\mul_1=\cos\thl_1$ is related 
to $\mul=\cos\thl$ by equation~(\ref{eq:ray}).
Equation~(\ref{eq:formal0}) can be rewritten in terms of $\IN(\muc,r)$ and 
$\SN(\muc,r)$ using the transformations $\INlab=\D^3\IN$ and $\SNlab=\D^3\SN$.
For outflows with $\beta\rightarrow 1$ this gives
\be
\label{eq:formal}
   \IN(\muc,r)=\int_{0}^\infty \SN(\muc_1,r_1)\,
    \left(\frac{1+\muc_1}{1+\muc}\right)^3
     \exp\left[-\taur(r_1,\muc_1\symb r)\right]\,d\taur(r_1,\muc_1\symb r).
\ee
Here one can substitute equation~(\ref{eq:tau3}) for 
$\taur(r_1,\muc_1\symb r)$. The identity 
$\taur(r_1,\muc_1\symb r)=-\taur(r,\muc\symb r_1)$ simplifies the integral. 

It is sufficient to know the source function $\SN(\muc,r)$ to 
reconstruct the solution for $\IN(\muc,r)$. In the model with isotropic
scattering the source function $\SN=\IN_0(r)$ does not depend on $\muc$.
Our numerical result for $\IN_0(r)$ agrees within a few percent with the 
following formula,
\be
\label{eq:SN}
    r^2\,\IN_0(r)=\KK\,\left\{\frac{3}{2}+\frac{1}{\pi}\arctan\left[
    \frac{1}{3}\left(\frac{\Rph}{r}-\frac{r}{\Rph}\right)\right]\right\},
\ee
where $\KK$ is a constant. It can be expressed in terms of the photon flux 
in the lab frame $\FNlab$ (\Sec~2.3), which satisfies $r^2\FNlab(r)=const$.
At $r/\Rph\gg 1$ we have $\IN_1=\IN_0$ and $\FNlab=4\pi\g(1+\beta)\IN_0$. 
This gives (with $\beta\rightarrow 1$)
\be
  \KK=\frac{r^2\FNlab(r)}{8\pi\g}.
\ee 
Substitution of equation~(\ref{eq:SN}) to equation~(\ref{eq:formal}) and its
integration over radius recovers $\IN(\muc,r)$ that was found by direct 
numerical integration of the transfer equation.

Similarly, the formal solution for transfer of energy is given by
 \be
\label{eq:formal_en}
   \I(\muc,r)=\int_{0}^\infty \S(\muc_1,r_1)\,
     \left(\frac{1+\muc_1}{1+\muc}\right)^4
     \exp\left[-\taur(r_1,\muc_1\symb r)\right]\,d\taur(r_1,\muc_1\symb r).
\ee
The model with isotropic coherent scattering has $\S(\muc,r)=\I_0(r)$. 
In this case, one can use the following formula,
\be
\label{eq:K}
   r^2\,\I_0(r)=K\,\left(\frac{1}{2}+\frac{\Rph}{r}\right)^{2/3},
\ee
where constant $K$ is determined by the inner boundary condition at 
$r\ll\Rph$. Equation~(\ref{eq:K}) remains approximately valid for cold 
outflows with Thomson scattering. For heated jets with significant 
Comptonization of radiation, $I_0(r)$ and $\S(\muc,r)$ are different and 
depend on the heating history.

In contrast, equation~(\ref{eq:SN}) remains an excellent approximation even 
for heated outflows, as long as scattering is the main source of opacity
and emissivity.

\subsection{Distribution of the last-scattering radius and angle}

Consider all photons escaping to infinity from a steady, spherically 
symmetric outflow. Let $\dN$ be the number of escaping photons 
per unit time, measured in the lab frame. One can think of $\dN$
as the rate of photon emission by a source distributed throughout 
the volume of the outflow and attenuated by the optical depth.
The photon emission rate from volume element $dV$ into solid angle 
$d\Omlab$ is $\jNlab\,dV\,d\Omlab$, where $\jNlab(\rls,\thlls)$
is the photon emissivity in the lab frame. It depends on the radial 
position of the emitter $dV$, $\rls$, and the emission angle with 
respect to the radial direction, $\thlls$.  The attenuation factor is 
$\exp[-\taur(\rls,\thlls\symb \infty)]$, which gives
\be
 \frac{d\dN}{dV\,d\Omlab}=\jNlab\,
        \exp\left[-\taur(\rls,\thlls\symb\infty)\right].
\ee
A similar expression holds for the angular distribution measured in the 
fluid frame (note that $\jNlab\,d\Omlab=\jN\,d\Om$ is invariant under 
Lorentz transformation). Substituting $dV=4\pi\rls^2\,d\rls$, 
$d\Om=d\phils\,d\mucls$ (where $\mucls=\cos\thcls$ describes the emission 
angle in the fluid frame), and integrating over $\phils$, we obtain 
\be
  \frac{d\dN}{d\rls d\mucls}=8\pi^2\rls^2\,\jN\,
          \exp\left[-\taur(\rls,\mucls\symb\infty)\right].
\ee
This equation describes the distribution of escaping photons over the 
last-scattering radius and angle.
The distribution can be normalized to unity if we divide it by 
$\dN=4\pi \rls^2\FNlab(\rls)$ where $\FNlab$ is the photon flux in the lab 
frame. This gives the probability distribution for $\rls$ and $\mucls$,
\be
\label{eq:dist}
  \frac{dP}{d\rls\,d\mucls}=2\pi\,\kappa\,\frac{\SN}{\FNlab}\,
                      \exp\left[-\taur(\rls,\mucls\symb\infty)\right].
\ee
Here we used the relation $\jN=\kappa\SN$ in the fluid frame, where 
$\kappa=\sigma n$ is the scattering opacity. Equation~(\ref{eq:dist})
shows that the distribution of the last-scattering radius and angle 
is proportional to the source function in the fluid frame, $\SN$, which is 
determined by the transfer solution. For the isotropic-scattering model 
one should use $\SN=\IN_0(r)$. Substitution of equations~(\ref{eq:tau4}) 
and (\ref{eq:SN}) to equation~(\ref{eq:dist}) gives
\be
\label{eq:dist1}
  \frac{dP}{d\ln\rls\,d\mucls}=\frac{\Rph}{4\rls}
        \left\{\frac{3}{2}+\frac{1}{\pi}\arctan\left[\frac{1}{3}
     \left(\frac{\Rph}{\rls}-\frac{\rls}{\Rph}\right)\right]\right\}
       \,\exp\left[-\frac{\rls}{6\Rph}
      \left(3+\frac{1-\mucls}{1+\mucls}\right)\right].
\ee
The corresponding distribution of $\rls$ and $\mulls$ is given by
$dP/d\rls d\mulls=\D^2\,dP/d\rls d\mucls$.

%########################################################################

% \newpage

% \bibliographystyle{apj}

% \begin{thebibliography}{}


\begin{references}
                                                                           
% \bibitem[]{}
Abramowicz, M. A., Novikov, I. D., Paczy\'nski, B. 1991, ApJ, 369, 175

% \bibitem[]{}
Beloborodov, A. M. 1998, ApJ, 496, L105

% \bibitem[]{}
Beloborodov, A. M. 1999, MNRAS, 305, 181

% \bibitem[]{}
Beloborodov, A. M. 2010, MNRAS, 407, 1033

% \bibitem[]{}
Castor, J. I. 1972, ApJ, 178, 779

% \bibitem[]{}
Chandrasekhar, S. 1960, Radiative Transfer (New York: Dover) 

% \bibitem[]{}
Ghirlanda, G., Celotti, A., \& Ghisellini, G., 2003, A\&A, 406, 879

% \bibitem[]{}
% Giannios, D., \& Spruit, H.~C., 2007, A\&A, 469, 1
Giannios, D. 2006, A\&A, 457, 763

% \bibitem[]{}
Goodman, J. 1986, ApJ, 308, L47

% \bibitem[]{}
Ioka, K., Murase, K., Toma, K., Nagataki, S., \& Nakamura, T., 2007, 670, L77

% \bibitem[]{}
Mihalas, D. 1980, ApJ, 237, 574

% \bibitem[]{}
Paczy\'nski, B. 1986, ApJ, 308, L43

% \bibitem[]{}
Paczy\'nski, B. 1990, ApJ, 363, 218

% \bibitem[]{}
Pe'er, A. 2008, ApJ, 682, 463

% \bibitem[]{}
Pe'er, A., \& Waxman, E. 2005, ApJ, 628, 857

% \bibitem[]{}
Prokof'ev, V. A., 1962, Sov. Phys. Doklady, 6, 861

% \bibitem[]{}
% Rybicki, G. B., \& Lightman, A. P. 1979, Radiative Processes in Astrophysics 
% (New York: Wiley) 

% \bibitem[]{}
Ryde, F., et al. 2010, ApJ, 709, L172

% \bibitem[]{}
Sobolev, V. V. 1963, A Treatise on Radiative Transfer (Princeton: Van Nostrand) 

% \bibitem[]{}
Spruit, H. C., Daigne, F., \& Drenkhahn, G. 2001, A\&A, 369, 694

% \bibitem[]{}
Thompson, C. 1994, MNRAS, 270, 480

% \bibitem[]{}
Vurm, I., Beloborodov A. M., \& Poutanen 2011, submitted to ApJ
(arXiv:1104.0394)

% \bibitem[]{}
\reference{ }
Vurm, I., \& Poutanen, J. 2009, ApJ, 698, 293

% \end{thebibliography}
\end{references}
\end{document}